\documentclass[floatfix,twocolumn,aps,prd,showpacs,amsmath,amssymb]{revtex4}
\usepackage{epsfig}
\usepackage[utf8]{inputenc}
\usepackage{slashed}

\begin{document}

\title{Pseudo Quantum Electrodynamics and Chern-Simons theory Coupled to Two-dimensional Electrons}

\author{Gabriel C. Magalhães$^{1}$, Van S\'ergio Alves$^{1}$, E. C. Marino$^2$, Leandro O. Nascimento$^{3}$}
\affiliation{$^1$ Faculdade de F\'\i sica, Universidade Federal do Par\'a, Av.~Augusto Correa 01, Bel\'em PA, 66075-110, Brazil  \\
$^2$Instituto de F\'\i sica, Universidade Federal do Rio de Janeiro, C.P. 68528, Rio de Janeiro RJ, 21941-972, Brazil
\\
$^3$Faculdade de Ci\^encias Naturais, Universidade Federal do Par\'a, C.P. 68800-000, Breves, PA,  Brazil }

\date{\today}

\begin{abstract}

We study a nonlocal theory that combines both the Pseudo quantum electrodynamics (PQED) and Chern-Simons actions among two-dimensional electrons. In the static limit, we conclude that the competition of these two interactions yields a Coulomb potential with a screened electric charge given by $e^2/(1+\theta^2)$, where $\theta$ is the dimensionless Chern-Simons parameter. This could be useful for describing the substrate interaction with two-dimensional materials and the doping dependence of the dielectric constant in graphene. In the dynamical limit, we calculate the effective current-current action of the model considering Dirac electrons. We show that this resembles the electromagnetic and statistical interactions, but with two different overall constants, given by $e^2/(1+\theta^2)$ and $e^2\theta/(1+\theta^2)$. Therefore, the $\theta$-parameter does not provide a topological mass for the Gauge field in PQED, which is a relevant difference in comparison with quantum electrodynamics. Thereafter, we apply the one-loop perturbation theory in our model. Within this approach, we calculate the electron self-energy, the electron renormalized mass, the corrected gauge-field propagator, and the renormalized Fermi velocity for both high- and low-speed
limits, using the renormalization group. In particular, we obtain a maximum value of the renormalized mass for $\theta\approx 0.36$. This behavior is an important signature of the model and relations with doping control of band gap size are also discussed in the conclusions.

\end{abstract}


\maketitle

\section{Introduction}

Quantum electrodynamics (QED) is one of the most well-succeed theories in physics, in particular, because
of the remarkable comparison between the experimental data for the electron g-factor and its theoretical prediction. Because the matter action is given by the Dirac theory, hence, most of its results are connected to electrons in particle physics. Recently, nevertheless, quantum electrodynamics has been shown a very useful tool in the physics of planar materials. Indeed, the experimental realization of graphene \cite{grapexp} has been an important bridge between quantum field theory and condensed matter physics in the last decade. The main reason is because of the low-energy excitations that behave as massless Dirac particles, with an effective velocity $v_F \approx c/300$, called Fermi velocity. This is a small fraction of the light velocity $c$ \cite{review}. The dynamics of electrons in graphene generates an experimental setup for observing the Klein tunneling at reasonable energy scales, in comparison with the energy scale in particle physics \cite{Kat}. Beyond this large set of applications, new two-dimensional materials made that bridge even bigger than before. For instance, silicene, phosphorene, germanene, and the transition metal dichalcogenides \cite{2Dmaterials} have been described by the massive Dirac theory at low energies. In literature, most of the applications have been discussed within a free-particle picture. On the other hand, a few applications have been made with some quantum-electrodynamical approach, in particular, the Pseudo Quantum Electrodynamics (PQED) has attracted great attention. The whole description of interactions, including those due to microscopic or statistical interaction, is still an open question in the physics of these planar materials.

It has been shown that PQED is the correct theory for describing electronic interactions among particles in the plane. This is a nonlocal model generated by a dimensional reduction of QED, after constraining the electrons to the plane \cite{marino}. For the sake of consistency with quantum theory, it has been shown that PQED is unitary \cite{unitarity}. Furthermore, it yields the Coulomb potential $V(r)\propto 1/r$ in the static limit, which is a desirable feature for applications in the physics of two-dimensional materials. Indeed, aiming for applications in condensed matter physics, several results have been obtained from this approach, for example, dynamical mass generation \cite{CSBPQED}; interaction driven quantum valley Hall effect \cite{PRX2015}; quantum corrections of the electron $g$-factor \cite{gfactor}; Yukawa potential in the plane \cite{Yukawa}; and electron-hole pairing (excitons) in transition metal dichalcogenides \cite{TMDPQED}. Nevertheless, all of these applications neglect the interaction of the matter with the Chern-Simons action at tree level. This interaction has been related to the effective action of the matter current, driven by the Chern-Simons action, after integrating out the gauge field. This effective action has been coined statistical interaction in Ref.~\cite{marino}. The Chern-Simons model has an intrinsic parameter $\theta$, which yields a topological mass in the quantum electrodynamics in (2+1) dimensions (QED$_3$), without breaking gauge symmetry \cite{Dunne}. 

In general grounds, the Chern-Simons action may be written either as an abelian or nonabelian gauge field. Here, we shall keep our attention only to the abelian case. The induction of the Chern-Simons action in the plane has been derived through a dimensional reduction of the topological action, derived from the Maxwell field. This action is given by the contraction of the electromagnetic field and its dual field. From this approach, it follows that the full interactions in the plane should include the $\theta$-parameter \cite{marino}. Another peculiar feature of the Chern-Simons theory is its induction, in the two-rank representation of the Dirac matrices, from one-loop quantum corrections to the photon propagator in QED$_3$. In this case, the induction of the Chern-Simons action is a realization of the so-called parity anomaly \cite{PRX2015}.

The interaction of the matter current with the Chern-Simons action yields a field theory for the Hall effect \cite{Bernevig}. In this case, the $\theta$-parameter is related to the Hall conductivity at the level of motion equations. Furthermore, applications of the Chern-Simons action have been discussed in several phenomena, such as superconductivity at high temperatures; fractional quantum Hall effect; anyons (particles with fractional statistics); and the Aharanov-Bohn effect \cite{Dunne}.

Given the relevance of both PQED and the Chern-Simons action for planar systems, it is relevant to describe the competition between the $\theta$-parameter and the electromagnetic coupling constant, namely, the fine-structure constant $\alpha$. In this work, we show that the Chern-Simons parameter increases the dielectric constant, through the scale $\alpha\rightarrow \alpha/(1+\theta^2)$, in the static limit. Within the dynamical regime, we apply the one-loop perturbation theory in our model, where we obtain: the electron mass renormalization; and the renormalization of the Fermi velocity. This model also has been discussed in Ref.~\cite{RQEDCS} in the light of the Coleman-Hill theorem.

The outline of this paper is the following: In Sec. II, we show the effective action for both, electromagnetic and statistical interactions. In Sec. III, we show the model, PQED with the Chern-Simons term. In Sec. IV, we calculate the static potential. In Sec. V, we calculate the electron self-energy for the isotropic case in one loop of perturbation theory. In Sec. VI, we calculate the renormalization of the Fermi velocity for high and low speeds using the renormalization group method. In Sec. VII, we review the main results obtained in this paper. We also include three Appendices. In App.~A, we discuss the Maxwell-Chern-Simons (MCS) theory. In App.~B, we calculate the renormalized mass and in App.~C we calculate the beta functions through the renormalization group for our model. 

\section{Effective action for both Electromagnetic and Statistical Interactions} 

In this section, we calculate both the electromagnetic and statistical interactions in the plane. In order to do so, let us consider the Euclidean version of the Chern-Simons theory in (2+1) dimensions, namely,
\begin{equation}\label{CS1}
{\cal L}_{{\rm CS}}=i\frac{\bar \theta}{2} \epsilon_{\mu\nu\alpha}{\cal {A}}_\mu \partial_\nu {\cal {A}}_\alpha+\bar e j_\mu {\cal {A}}_\mu+\frac{\lambda \bar \theta}{2}\frac{{\cal {A}}^\mu \partial_\mu \partial_\nu {\cal {A}}^\nu}{(-\Box)^{1/2}} ,
\end{equation}
where $\bar \theta$ is the Chern-Simons parameter with mass dimension ($[\bar \theta]=1$), ${\cal {A}}_\mu$ is the statistical field, $\Box$ is the d'Alembertian operator, $\bar e$ is the electric charge ($[\bar e]=1/2$), and $j_\mu$ is the matter current. The last term is included for calculating the gauge-field propagator. Eq.~(\ref{CS1}) yields an effective description for the Hall current at the level of motion equation \cite{marino}. Furthermore, it describes particles with an arbitrary statistics, the so-called Anyons. In this case, the arbitrary statistics emerges due to a particular Gauge transformation in the matter field. The composite field binds with an magnetic-like flux and, after interchanging the position of two particles, the two-particle wavefunction acquires a phase given by $e^{i \pi \Delta s}$, where the parameter $\Delta s= \bar e^2/(2\pi \bar\theta)$ describes the statistics of the matter field \cite{CSST}. 

Integration over ${\cal {A}}_\mu$ in Eq.~(\ref{CS1}) yields the effective action for the current matter $j_\mu$, i.e,
\begin{equation}
{\cal L}^{\rm{CS}}_{\rm{eff}.}[j_\mu]=i \frac{\bar e^2}{2\bar \theta} j^\mu(x)\epsilon_{\mu\nu\alpha}\left[\frac{\partial^\alpha}{2(-\Box)}\right]_{(x-y)} j^\nu(y), \label{CS2}
\end{equation}
which represents the statistical interaction \cite{marino}, obtained after we use the conservation of the matter current. Next, we review the effective electromagnetic interaction in the plane.

The PQED action reads
\begin{equation}
{\cal L}_{{\rm 3D}}=\frac{1}{2}\bar F^{\mu\nu} (-\Box)^{-1/2} \bar F_{\mu\nu}+\frac{\lambda}{2}\frac{\bar A^\mu \partial_\mu \partial_\nu \bar A^\nu}{(-\Box)^{1/2}}+e j^\mu \bar A_\mu, \label{PQED0}
\end{equation}
where $\bar F_{\mu\nu}=\partial_\mu \bar A_\nu-\partial_\nu \bar A_\mu$ with $\bar A_\mu$ being the pseudo-electromagnetic field. Note the $\bar A_\mu$ is not the real electromagnetic field that lives in (3+1)D. The action of PQED is completely defined in (2+1)D, hence, $\bar A_\mu$ lives in the plane. In particular, electrons interact trough the Coulomb potential in the static limit. Similarly to the previous calculation, we find 
\begin{equation}
{\cal L}^{\rm{PQED}}_{\rm{eff}.}[j_\mu]=-\frac{e^2}{2}j^\mu(x) \left[\frac{1}{2(-\Box)^{1/2}}\right]_{(x-y)}j_\mu(y). \label{PQED01}
\end{equation}
Eq.~(\ref{PQED01}) is the same effective action generated by quantum electrodynamics with particles confined to the plane \cite{marino}. In fact, this is the origin of the square root of $\Box$ in PQED. Here, our main goal is to obtain a model that describes at the same time both Eq.~(\ref{CS2}) and Eq.~(\ref{PQED01}).
Perhaps, the first attempt would be to consider the combination of quantum electrodynamics in (2+1) dimensions and the Chern-Simons theory coupled to the matter field. The resulting effective action, however, does not correctly describe either the electronic or the statistical interactions. We shall discuss this attempt in App.~A. 

The desired theory has been found in Ref.~\cite{marino}, where it is shown that
\begin{eqnarray}
{\cal L}[{\cal A}_\mu,\bar A_\mu]&=&\frac{1}{2}\bar F^{\mu\nu} (-\Box)^{-1/2} \bar F_{\mu\nu}+\frac{\lambda}{2}\frac{\bar A^\mu \partial_\mu \partial_\nu \bar A^\nu}{(-\Box)^{1/2}}\nonumber\\
&+&e j^\mu \bar A_\mu+i \frac{\bar \theta}{2} \epsilon_{\mu\nu\alpha}{\cal A}^\mu \partial^\nu {\cal A}^\alpha+ \bar e j^\mu {\cal A}_\mu \label{2FM}
\end{eqnarray}
generates the statistical interaction, in Eq.~(\ref{CS2}), and the electromagnetic interaction, given by Eq.~(\ref{PQED01}). Indeed, integration over $\bar A_\mu$ and ${\cal A}_\mu$  in Eq.~(\ref{2FM}) yields the sum of these interactions. Note that this model interpolates two Gauge fields, which have different canonical dimensions whether $[\bar\theta]=1$ and $[\bar e]=1/2$ (as it necessarily happens in MCS theory). This, nevertheless, may be circumvented by assuming that $\bar e$ and $\bar\theta$ are dimensionless constants. In this case, it is straightforward that $[\bar A_\mu]=[{\cal A}_\mu]$, with $[\bar e]=[e]$ as it is expected. Next, we propose an alternative approach for describing these basic interactions by taking the following \textit{ansatz} $\bar A_\mu={\cal A}_\mu$ for the Gauge fields in Eq.~(\ref{2FM}). A similar approach has been used in Ref.~\cite{Kondo} for investigating  dynamical mass generation in MCS theory. Recently, this condition also has been used for studying the parity anomaly in MCS theory at the lattice \cite{anomaly}.

\section{The PQED-Chern-Simons Model}
We start with the action, in the Euclidean space, given by
\begin{eqnarray}
{\cal L}&=&\frac{1}{2}F^{\mu\nu} (-\Box)^{-1/2}F_{\mu\nu}+\frac{\lambda}{2} A^\mu \partial_\mu (-\Box)^{-1/2}\partial_\nu A^\nu \nonumber\\
&+&{\cal L}_M[\psi]+i \frac{\theta}{2} \epsilon_{\mu\nu\alpha}A^\mu \partial^\nu A^\alpha+e j^\mu A_\mu,  \label{action}
\end{eqnarray}
where $e$ is the electric charge, $j_\mu=\bar\psi\gamma_\mu\psi$ is the matter current, ${\cal L}_M[\psi]$ is the Dirac action 
for the matter field, explicitly $\bar{\psi}(i\slashed{\partial}-m)\psi$, and $\lambda$ is a gauge-fixing parameter. $A_\mu$ is a Gauge field which we call PQED-Chern-Simons field and $\theta$ is a dimensionless parameter that resembles $\bar\theta$ in Eq.~(\ref{CS1}). Eq.~(\ref{action}) is meant to describe the coupling of PQED with the Chern-Simons action. Note that Eq.~(\ref{action}) is obtained from Eq.~(\ref{2FM}) after we consider $(\bar A_\mu, {\cal A}_\mu)\rightarrow A_{\mu}$ and $\bar e=0$ without loss of generality (the minimal coupling is already given by $e j^\mu A_\mu$).

Despite taking a simplest version of Eq.~(\ref{2FM}), our model also generates the current-current correlation function obtained in Sec.~II. Indeed, after integrating out $A_\mu$ in Eq.~(\ref{action}), we find
\begin{equation}
{\cal L}_{\rm{eff.}}[j_\mu]=-\frac{e^2}{2}j^\mu \Delta^{(0)}_{\mu\nu}j^\nu. \label{eff}
\end{equation}
Note that the kind of ${\cal L}_M[\psi]$ is needless for calculating Eq.~(\ref{eff}). $\Delta^{(0)}_{\mu\nu}$ is the gauge-field propagator, given by
\begin{equation}\label{gaugeprop}
\begin{split}
\Delta^{(0)}_{\mu\nu}&=\frac{\delta_{\mu\nu}}{2(-\Box)^{1/2}(1+\theta^2)}-\frac{i\theta\epsilon_{\mu\nu\alpha}\partial^\alpha}{2(-\Box)(1+\theta^2)}+\\
&+\frac{1}{\lambda(-\Box)^{1/2}}\frac{\partial_{\mu}\partial_{\nu}}{\Box}.
\end{split}
\end{equation}
Next, we use Eq.~(\ref{gaugeprop}) in Eq.~(\ref{eff}), yielding
\begin{eqnarray}\label{eff2}
{\cal L}_{\rm{eff.}}[j_\mu]&=&-\frac{e^2}{2(1+\theta^2)}j^\mu \left[\frac{1}{2(-\Box)^{1/2}}\right]j_\mu\nonumber \\
&+&\frac{ie^2\theta}{2(1+\theta^2)}j^\mu\epsilon_{\mu\nu\alpha}\left[\frac{\partial^\alpha}{2(-\Box)}\right] j^\nu. 
\end{eqnarray}
Comparison with Eq.~(\ref{CS2}) and Eq.~(\ref{PQED01}) shows that the first term on the right-hand side of Eq.~(\ref{eff2}) describes the electromagnetic interaction. Actually, for the sake of accuracy, we should consider a screening effect for the electric charge, namely, $e^2/(1+\theta^2)\rightarrow e^2$ in order to obtain an exact correspondence. On the other hand, the second term is very similar to the statistical interaction we have discussed in Sec.~II. Eq.~(\ref{eff2}) also ensures that our action provides the same (after properly scaling the parameters $e$ and $\theta$) current-current correlation functions calculated from Eq.~(\ref{2FM}). This is relevant for calculating physical observables, for example, the dc electric conductivity, in the light of the Kubo formula \cite{Bernevig}, for PQED applied to graphene \cite{PRX2015}. Thereafter, we shall prove that the effects of the Chern-Simons action in PQED is connected to screening effects instead of provide a topological mass as it happens in MCS model \cite{Kondo}. Here, it is relevant to comment that the Chern-Simons action also may be obtained through a dimensional projection, similarly to what has been done for PQED. In this case, the corresponding theory in (3+1)D is given by a total derivative of a topological current \cite{marino}. Furthermore, the model in Eq.~(\ref{action}) also has been obtained through the Bosonization of massless Dirac fermions in (2+1)D at one-loop approximation \cite{marinolivro}. 

The Feynman rules for this model are the gauge-field propagator, in momentum space, given by
\begin{equation}\label{PMCSgaugepropagatorwhithoutGF}
\Delta^{(0)}_{\mu\nu}(p)=\frac{\delta_{\mu\nu}}{2(p^{2})^{1/2}(1+\theta^{2})}+\frac{\theta\epsilon_{\mu\nu\alpha}p^{\alpha}}{2p^{2}(1+\theta^{2})}+\frac{1}{\lambda\sqrt{p^{2}}}\frac{p_{\mu}p_{\nu}}{p^{2}}.
\end{equation}
In our calculations we shall use the Landau gauge ($\lambda=\infty$). The electron-field propagator is
\begin{equation}\label{fermionprop}
S^{(0)}_{F}=\frac{-1}{\gamma^\mu p_\mu-m}
\end{equation}
and
\begin{equation}\label{vertex}
\Gamma^{\mu}=e\gamma^{\mu}
\end{equation}
is the vertex interation. In the next section, we calculate the static potential for this model. The gamma matrix are rank 4 with $(\gamma^0)^2=(\gamma^1)^2=(\gamma^2)^2=-1$, satisfying $\left\lbrace\gamma^{\mu},\gamma^{\nu}\right\rbrace=-2\delta^{\mu\nu}$.

\section{Static interaction}

The Fourier transform of the static propagator of gauge field given by Eq.~(\ref{PMCSgaugepropagatorwhithoutGF}), basically $\Delta^{(0)}_{00}(p_0=0,{\bf p})$, generates the static potential $V(r)$, therefore,
\begin{equation}\label{potencial1}
V(r)=e^{2}\int\dfrac{d^{2}p}{(2\pi)^{2}}\frac{\exp{\left(-i\vec{p}\cdot\vec{r}\right)}}{2(\vert\vec{p}\vert^{2})^{1/2}(1+\theta^{2})}.
\end{equation}
After solving the integration over $p$ \cite{gradstheyn}, we have
\begin{equation}\label{potencial4}
V(r)=\frac{e^{2}}{4\pi(1+\theta^{2})}\frac{1}{\vert\vec{r}\vert}.
\end{equation}
For $\theta=0$, we have the Coulomb potential, but for $\theta\neq 0$, we have an overall factor that behaves as an effective electric susceptibility. 


\begin{figure}[hbtp]
\centering
\hspace{-.25cm}\includegraphics[scale=0.7]{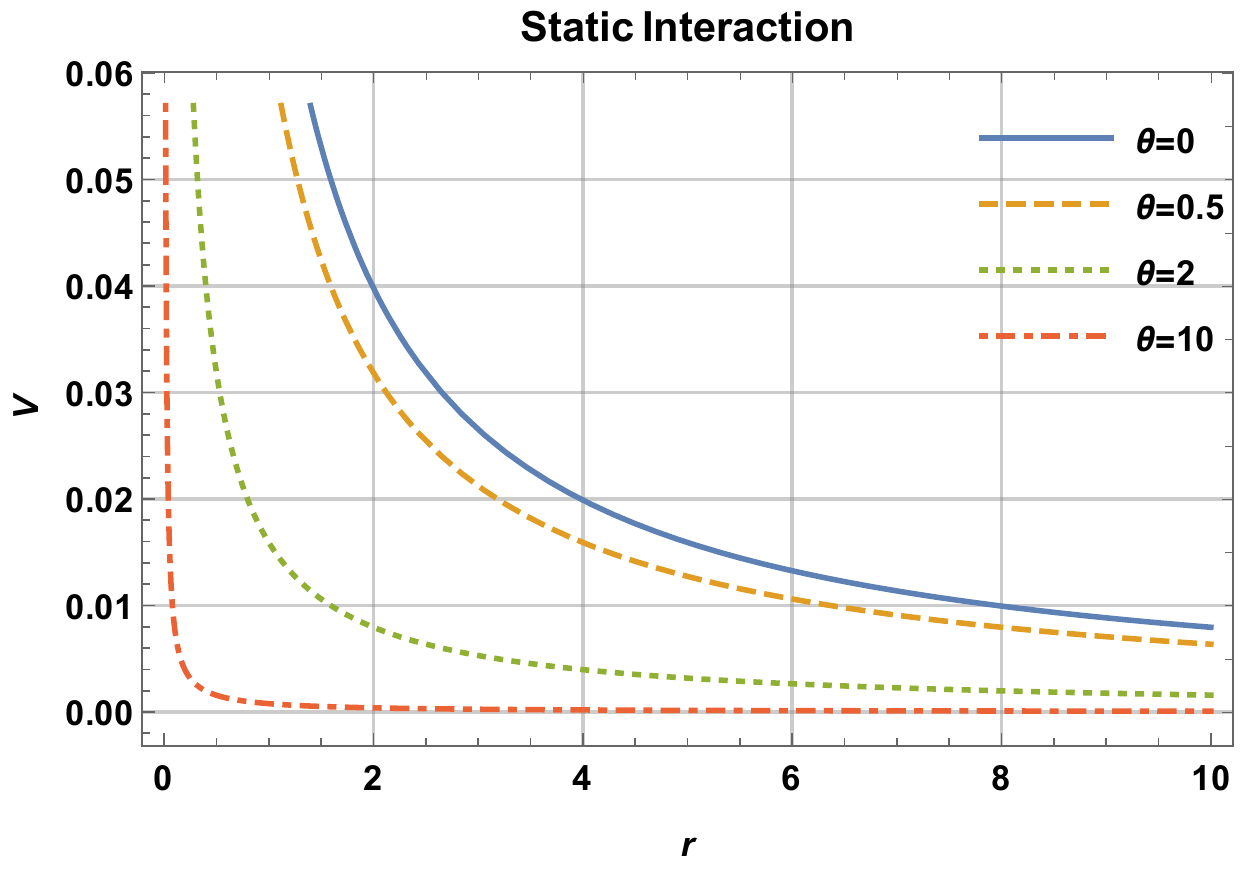}
\caption{\textit{Static potential of PQED-Chern-Simons theory}. This plot shows the behavior of the static potential Eq.~(\ref{potencial4}) as a function of the distance between the fermions, varying the parameter $\theta$. The solid line (blue), dashed (orange), dotted (green) and dot-dashed (red) show the static potential as a function of the position for $\theta=0$, $\theta=0.5$, $\theta=2$ and $\theta=10$, respectively.}
\label{grafpoten}
\end{figure}


In Fig.~\ref{grafpoten}, we plot this potential for some values of $\theta$. We conclude that for very large values of $\theta$, the electron-electron interaction becomes very weak in comparison with the unscreened Coulomb potential. This effect only occurs, within our model, because $\theta$ is dimensionless.

\section{The isotropic self-energies}

\subsection{The fermion self-energy and the mass renormalization}

The electron self-energy is given by
\begin{equation}\label{split}
\Sigma(p,m)=e^{2}\int\frac{d^{D}k}{(2\pi)^{D}}\gamma^{\mu}S_{F}(p-k)\gamma^{\nu}\Delta^{(0)}_{\mu\nu}(k),
\end{equation}
where $D$ is the space-time dimension.

After using Eq.~(\ref{PMCSgaugepropagatorwhithoutGF}) in Eq.~(\ref{split}), we split the electron self-energy in two terms, namely

\begin{equation}
\label{sigma}
\Sigma (p) = \frac{1}{(1+\theta^{2})}\left[\Sigma^{\textrm{PQED}}_{\textrm{ISO}}(p)+\Sigma_{\textrm{CS}}(p)\right],
\end{equation}
where

\begin{equation}
\label{sigma1}
\Sigma^{\textrm{PQED}}_{\textrm{ISO}}(p)=e^{2}\mu^{\epsilon}\int\frac{d^{D}k}{(2\pi)^{D}} \gamma^{\mu}S_{\textrm{F}}(p-k)\gamma^{\nu}\Delta^{\textrm{PQED}}_{\mu\nu}(k),
\end{equation}
with
\begin{equation}
\Delta^{\textrm{PQED}}_{\mu\nu}(k)=\frac{\delta_{\mu\nu}}{2\sqrt{k^{2}}},
\end{equation}
and
\begin{equation}\label{Sigma}
\begin{split}
\Sigma_{\textrm{CS}} (p)&=-\frac{\theta}{2} e^{2}\mu^{\epsilon}\epsilon_{\mu\alpha\nu}\times\\
&\int\frac{d^{D}k}{(2\pi)^{D}}\frac{(\gamma^{\mu}\slashed{p}\gamma^{\nu}-\gamma^{\mu}\slashed{k}\gamma^{\nu}+m\gamma^{\mu}\gamma^{\nu})}{[(p-k)^{2}+m^{2}]}\frac{k^{\alpha}}{k^{2}}.
\end{split}
\end{equation}

Eqs.(\ref{sigma1}) and (\ref{Sigma}) represent the effects of the isotropic PQED and CS term for the electron self-energy, respectively. We have also introduced the scale parameter $\mu$ by replacing $e\rightarrow e \mu^{\epsilon/2}$, where $\epsilon = 3-D$ is the dimensional regulator.

For calculating $\Sigma^{\textrm{PQED}}_{\textrm{ISO}}$ we have used the Feynman's trick,
\begin{equation}
\frac{1}{ab^{1/2}}=\frac{1}{2}\int^{1}_{0}dx\frac{(1-x)^{-1/2}}{[ax+b(1-x)]^{3/2}},
\end{equation}
where $x$ is named Feynman's parameter and the constants $a$ and $b$ are chosen as $a=(p-k)^{2}+m^{2}$ and $b=k^{2}$, respectively. Thereafter, using the dimensional regularization scheme \cite{Bollini}, we find the isotropic electron self-energy for PQED,
\begin{equation}
\begin{split}
\Sigma^{\textrm{PQED}}_{\textrm{ISO}}(p)&=\frac{e^{2}}{8\pi^{2}}\int_{0}^{1}dx(1-x)^{-1/2}\left(5(1-x)\slashed{p}-3m\right)\frac{1}{\epsilon}+\\
&\hspace{.4cm}-\frac{e^{2}}{16\pi^{2}}\int_{0}^{1}dx(1-x)^{-1/2}\cdot\\
&\hspace{.4cm}\cdot\left[\left(2+5\gamma-5\ln\left(\frac{4\pi\mu^{2}}{\delta}\right)\right)(1-x)\slashed{p}+\right.\\
&\hspace{.4cm}\left. +\left(3\ln\left(\frac{4\pi\mu^{2}}{\delta}\right)-3\gamma-2\right)m\right],
\end{split}
\end{equation}
such that $\delta=xm^{2}+x(1-x)p^{2}$ and $\gamma$ is the Euler - Mascheroni constant ($\sim 0.577$).
 
The second term of Eq.(\ref{sigma}) can be written as

\begin{equation}
\begin{split}
\bar{\Sigma}_{\textrm{CS}}(p)&=\frac{1}{1+\theta^2}\Sigma_{\textrm{CS}}(p)-\frac{\theta e^{2}\mu^{\epsilon}\epsilon_{\mu\alpha\nu}}{2(1+\theta^{2})}\times\\
&\int\frac{d^{D}k}{(2\pi)^{D}}\frac{(\gamma^{\mu}\slashed{p}\gamma^{\nu}-\gamma^{\mu}\slashed{k}\gamma^{\nu}+m\gamma^{\mu}\gamma^{\nu})}{[(p-k)^{2}+m^{2}]}\frac{k^{\alpha}}{k^{2}}.
\end{split}
\end{equation}

Using the following relation
\begin{equation}\label{relation1}
\gamma^{\mu}\gamma^{\nu}\gamma^{\alpha}=\epsilon^{\mu\nu\alpha}-g^{\mu\nu}\gamma^{\alpha}-g^{\nu\alpha}\gamma^{\mu}+g^{\mu\alpha}\gamma^{\nu},
\end{equation}
we find
\begin{equation}\label{autoiso2}
\bar{\Sigma}_{\textrm{CS}}(p)=\frac{\theta e^{2}\mu^{\epsilon}}{2(1+\theta^{2})}\int\frac{d^{D}k}{(2\pi)^{D}}\frac{(p\cdot k - k^{2} + m\slashed{k})}{[(p-k)^{2}+m^{2}]k^{2}}.
\end{equation}
Next, using another Feynman parametrization,
\begin{equation}
\frac{1}{ab}=\int^{1}_{0}dy\frac{1}{[ax+b(1-x)]^{2}},
\end{equation}
we may rewrite $\bar{\Sigma}_{\textrm{CS}}$ as
\begin{equation}
\begin{split}
\bar{\Sigma}_{\textrm{CS}}(p)&=\frac{\theta e^{2}\mu^{\epsilon}}{2(1+\theta^{2})}\times\\
&\int_{0}^{1}dx\int\frac{d^{D}k}{(2\pi)^{D}}\frac{x(1-x)p^{2} +xm\slashed{p} -k^{2}}{[k^{2}+\Delta]^{2}},
\end{split}
\end{equation}
where $\Delta=xm^{2}+x(1-x)p^{2}$. We apply the dimensional regularization scheme for calculating the loop integral, hence,
\begin{equation}
\begin{split}
\bar{\Sigma}_{\textrm{CS}}(p)&=\frac{\theta e^{2}}{16\pi(1+\theta^{2})}\int_{0}^{1}dx\frac{4x(1-x)p^{2}+xm(\slashed{p}+3m)}{[xm^{2}+x(1-x)p^{2}]^{\frac{1}{2}}}\\
&=\frac{\theta e^{2}}{16\pi(1+\theta^{2})}\left\lbrace Am+B\slashed{p}\right\rbrace ,
\end{split}
\end{equation}
where
\begin{equation}
\begin{split}
A=\frac{1}{m}\int_{0}^{1}dx\frac{4x(1-x)p^{2}+3xm^{2}}{[xm^{2}+x(1-x)p^{2}]^{\frac{1}{2}}}
\end{split}
\end{equation}
and
\begin{equation}
\begin{split}
B=\int_{0}^{1}dx\frac{xm}{[xm^{2}+x(1-x)p^{2}]^{\frac{1}{2}}}.
\end{split}
\end{equation}
To obtain a finite amplitude, we use the minimal subtraction method, which consist of introducing counter-terms ($CT$) in Eq.~(\ref{Sigma}). Therefore, the renormalized electron self-energy reads 
\begin{equation}
\Sigma^R=\lim_{\mu,\epsilon \rightarrow 0} (\Sigma(p,\mu,\epsilon)-CT)= C\slashed{p}+Dm, \label{sigR}
\end{equation}
where the coefficients $C$ and $D$ are expressed by
\begin{equation}\label{C}
\begin{split}
C&=-\frac{e^{2}}{16\pi^{2}(1+\theta^{2})}\times\\
&\int_{0}^{1}dx\left\lbrace\left[2+5\gamma+5\ln\left(\frac{\delta}{m^{2}}\right)\right]\sqrt{1-x}\right.+\\
&-\left.\frac{xm\pi\theta}{[xm^{2}+x(1-x)p^{2}]^{\frac{1}{2}}}\right\rbrace
\end{split}
\end{equation}
and
\begin{equation}\label{D}
\begin{split}
D&=\frac{e^{2}}{16\pi^{2}(1+\theta^{2})}\int_{0}^{1}dx\left\lbrace\frac{\left[3\gamma+2+3\ln\left(\frac{\delta}{m^{2}}\right)\right]}{\sqrt{1-x}}\right.+\\
&+\left.\frac{\pi\theta\left[4x(1-x)p^{2}+3xm^{2}\right]}{m[xm^{2}+x(1-x)p^{2}]^{\frac{1}{2}}}\right\rbrace.
\end{split}
\end{equation}

\begin{figure}
\centering
\hspace{-.25cm}\includegraphics[scale=.7]{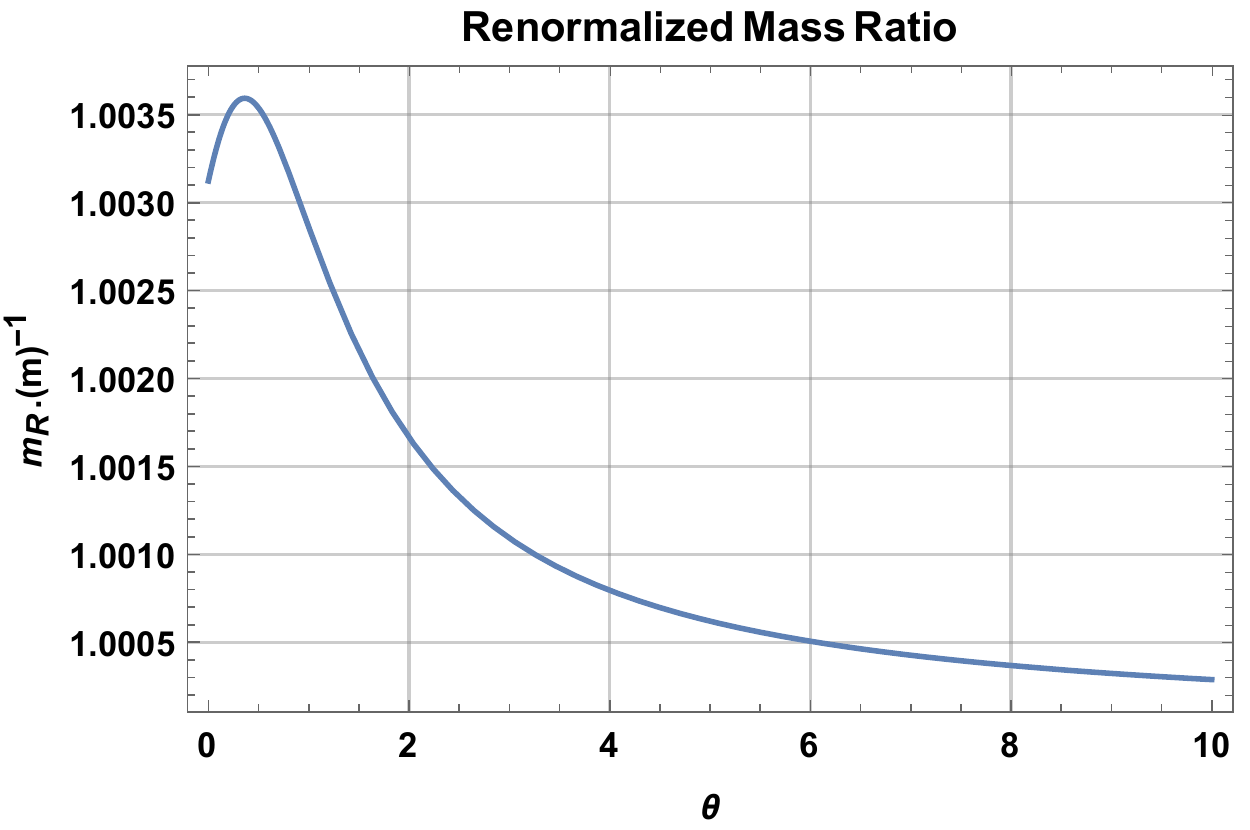}
\caption{\textit{The variation of renormalized mass ratio with $\theta$}. This plot shows the behavior of the variation of renormalized mass ratio Eq.~(\ref{mrm-1}) as a function of $\theta$, considering $e^{2}/4\pi=1/137$.}\label{mrgraf}
\end{figure}

The renormalized mass $m_{\textrm{R}}$ is given by the pole of the corrected propagator. Similarly  to the calculation that has been done in \cite{Nascimento},
we found that the renormalized mass ratio is  (for more details see Appendix B)
\begin{equation}\label{mrm-1}
\begin{split}
\frac{m_{\textrm{R}}}{m}&=1+C+D,\\
&=1-\frac{\alpha^{\textrm{ISO}}_{\textrm{eff}}}{4\pi}\int_{0}^{1}dx\left\lbrace \frac{[\gamma+2\ln(x)](2-5x)-2x}{\sqrt{1-x}}+\right.\\
&-\left.\frac{4x^{2}\pi\theta}{\sqrt{2x-x^{2}}}\right\rbrace ,
\end{split}
\end{equation}
where $m$ is the bare mass of the electron and $\alpha^{\textrm{ISO}}_{\textrm{eff}}=e^{2}/4\pi(1+\theta^{2})$.
Graphically, the renormalized mass ratio is shown in Fig.~\ref{mrgraf}, where the global maximum value of renormalized mass was obtained at $\theta\cong 0.36$. Furthermore, for large values of $\theta$ the mass does not renormalize.


\subsection{The photon self-energy and screened static interaction}

From now on, let us assume that the matter field is massless. Hence, the full propagator of the gauge field, considering the insertion of the photon self-energy \cite{Appel,CSBPQED}, is given by
\begin{equation}\label{fullG}
\Delta^{\textrm{Full}}_{\mu\nu}(p)=\frac{\sqrt{p^{2}}\left[2(1+\theta^{2})+\frac{e^{2}}{8}\right]P_{\mu\nu}+2\theta (1+\theta^{2})\epsilon_{\mu\nu\lambda}p^{\lambda}}{p^{2}\left[2(1+\theta^{2})+\frac{e^{2}}{8}\right]^{2}},
\end{equation}
where $P_{\mu\nu}$ is called the transverse operator, given by $P_{\mu\nu}=\delta_{\mu\nu}-p_{\mu}p_{\nu}/p^{2}$.
Note that we have used the four-rank representation of the Dirac matrices in Eq.~(\ref{fullG}).

In the static limit, within the Landau gauge, the gauge-field propagator reads
\begin{equation}
\Delta^{\textrm{Full}}_{00}(\textbf{p})=\frac{1}{2\sqrt{\textbf{p}}\left( 1+\theta^{2}+\frac{e^{2}}{16}\right)}.
\end{equation}
After solving the Fourier transform of the static propagator, we find the potential
\begin{equation}\label{staticfull}
V(r)=\frac{e^{2}}{4\pi\left(1+\theta^{2}+\frac{e^{2}}{16}\right)}\frac{1}{\vert\vec{r}\vert}.
\end{equation}
Eq.~(\ref{staticfull}) yields the total screening effect of the potential. Therefore, we have shown that the whole description of the screening depends on both the statistical parameter $\theta$ and the quantum corrections given by the factor $e^2/16$. 

In the next section, we calculate the anisotropic self-energy and the Fermi velocity renormalization for high and low-speed regimes.

\section{The anisotropic self-energy and the Fermi velocity renormalization}
Let us consider a soft breaking of the Lorentz symmetry \cite{mgomes} in the massless fermionic sector given by the following Lagrangean
\begin{equation}\label{action2}
\begin{split}
{\cal L}&=\frac{1}{2}F^{\mu\nu} (-\Box)^{-1/2}F_{\mu\nu}+\frac{\lambda}{2} A^\mu \partial_\mu (-\Box)^{-1/2}\partial_\nu A^\nu+ \\
&+\bar{\psi}(i\gamma^{0}\partial_{0}+iv_{\textrm{F}}\gamma^{i}\partial_{i})\psi+i \frac{\theta}{2} \epsilon_{\mu\nu\alpha}A^\mu \partial^\nu A^\alpha+e j^\mu A_\mu. 
\end{split}
\end{equation}
Note that in the model of Eq.~(\ref{action2}) a possible mass term, such as $m\bar\psi\psi$, does not changes the renormalization of $v_F$ \cite{luis}.  

From Eq.~(\ref{action2}), we conclude that the fermion propagator is given by
\begin{equation}
S^{(0)}_{F}(p)=\frac{\gamma^{\mu}\bar{p}_{\mu}}{\bar{p}^2},
\end{equation}
where $\bar{p}^{\mu}=(p_0,v_{\textrm{F}}\bf{p})$, and $\bar{p}^2=p_0^2+v_{\textrm{F}}^2\bf{p}^2$. On the other hand, the photon propagator reads
\begin{equation}\label{photin anisotropic}
\Delta^{(0)}_{\mu\nu}(p)=\frac{\delta_{\mu\nu}}{2(p^{2})^{1/2}(1+\theta^{2})}+\frac{i\theta\epsilon_{\mu\nu\alpha}p^{\alpha}}{2p^{2}(1+\theta^{2})},
\end{equation}
where we have adopted the Landau gauge ($\lambda = \infty$), $p^{\mu}=(p_0,\bf{p})$ with $p^2=p_0^2+\bf{p}^2$, and  vertex structure of Eq.~(\ref{vertex}) becomes
\begin{equation}\label{vertex2}
\Gamma^{\mu}=e\left(\gamma^{0},v_{\textrm{F}}\gamma^{i}\right).
\end{equation}
Note that in this section we consider $c=1$ for the sake of simplicity.

The quantum corrections to $v_F$ are obtained from the electron self-energy, which is given by
\begin{equation}\label{eq36}
\Sigma(p)=\int\frac{d^{D}k}{(2\pi)^{D}}\Gamma^{\mu}S_{F}(\bar{p}-\bar{k})\Gamma^{\nu}\Delta_{\mu\nu}(k).
\end{equation}
From Eq.~(\ref{eq36}), it is possible to split the electron self-energy into two terms. The first one contains the effects of PQED, with the overall factor $1/(1+\theta^{2})$, and the second one is the additional Chern-Simons contribution. Hence,
\begin{equation}\label{Sigma2}
\begin{split}
\Sigma(p) &= \frac{\Sigma^{\textrm{PQED}}_{\textrm{ANI}}}{(1+\theta^{2})}+\frac{\theta e^{2}}{(1+\theta^{2})}\times\\
& \times\epsilon_{\mu\alpha\nu}\int\frac{d^{D}k}{(2\pi)^{D}}\frac{(\gamma^{\mu}\slashed{\bar{p}}\gamma^{\nu}-\gamma^{\mu}\slashed{\bar{k}}\gamma^{\nu})}{[(\bar{p}-\bar{k})^{2}]}\frac{k^{\alpha}}{k^{2}}.
\end{split}
\end{equation}
In what follows, we will calculate these diagrams using the dimensional regularization procedure as a way to obtain finite Feynman amplitudes. First, we perform the integration over $k_{0}$ and, after that, the remaining integral over $\textbf{k}$. We use standard formulas of dimensional regularization in spatial dimension $d=2-\epsilon$, and that $e\rightarrow e{\mu}^{\epsilon /2}$, which keeps the correct dimensionality of the integrals.

Note that the integrals in the second term of Eq.~(\ref{Sigma2}) are finite. Hence, using the method of renormalization group (see App.~C), we conclude that this term does not contribute to the Fermi velocity renormalization. $\Sigma^{\textrm{PQED}}_{\textrm{ANI}}$ is the anisotropic electron self-energy for PQED, given by
\begin{equation}
\begin{split}
\Sigma^{\textrm{PQED}}_{\textrm{ANI}}(p)&=e^{2}\mu^{\epsilon}\int\frac{d^{D}k}{(2\pi)^{D}} \gamma^{\mu}S_{\textrm{F}}(\bar{p}-\bar{k})\gamma^{\nu}\Delta_{\mu\nu}(k)\\
&=\frac{e^{2}\mu^{\epsilon}}{8\pi^{2}}\left\lbrace -(1-2v_{\textrm{F}}^{2})\gamma^{0}p_{0}I_{1}+\right.\\
&\left.+v_{\textrm{F}}\gamma^{i}p_{i}I_{2}\right\rbrace\frac{1}{\epsilon}\\
&+\mbox{Finite\,\,Terms},
\end{split}
\end{equation}
where
\begin{equation}
\begin{split}
I_{1}&=\int_{0}^{1}dx\frac{(1-x)^{1/2}}{x(1-v_{\textrm{F}}^{2})-1},\\
I_{2}&=\int_{0}^{1}dx\frac{(1-x)^{-1/2}\left(1+\frac{xv_{\textrm{F}}^{2}}{x(1-v_{\textrm{F}}^{2})-1}\right)}{x(1-v_{\textrm{F}}^{2})-1}.
\end{split}
\end{equation}
Therefore, the running Fermi velocity $\beta_{v_{\textrm{F}}}$ may be written as
\begin{equation}\label{runningvf}
\begin{split}
\beta_{v_{\textrm{F}}}&=\frac{e^{2}}{8\pi^{2}(1+\theta^{2})}\left[v_{\textrm{F}} I_{2}+(1-2v_{\textrm{F}}^{2})I_{1}\right]\\
&=-\frac{e^{2}}{8\pi^{2}(1+\theta^{2})}\times\\
&\int_{0}^{1}dx\frac{\sqrt{1-x}}{1-x(1-v_{\textrm{F}}^{2})}\left[ 1-2v_{\textrm{F}}^{2}+\frac{1}{1-x(1-v_{\textrm{F}}^{2})}\right],
\end{split}
\end{equation}
where $\beta_{v_{\textrm{F}}}=\mu(\partial v^{\textrm{R}}_{\textrm{F}}/\partial\mu)$.
 
\begin{figure}[hbtp]
\centering
\hspace{-.25cm}\includegraphics[scale=0.7]{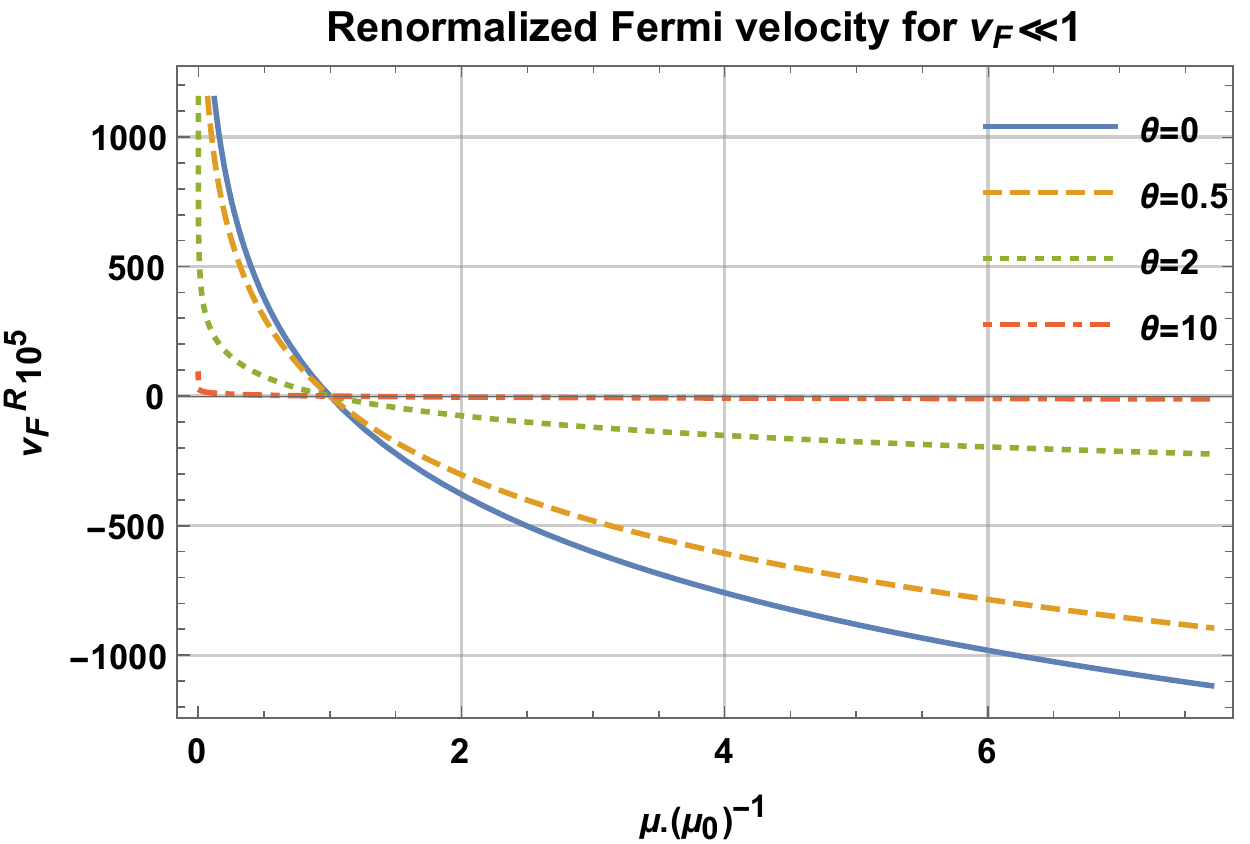}
\caption{\textit{The renormalized Fermi velocity for $v_{\textrm{F}}\ll 1$}. This plot shows the behavior of the renormalized Fermi velocity Eq.~(\ref{vfrstaticregime}) as a function of $\mu/\mu_{0}$, varying the parameter $\theta$, where the solid line (blue), dashed (orange), dotted (green) and dot-dashed (red) are for $\theta=0$, $\theta=0.5$, $\theta=2$ and $\theta=10$, respectively.}
\label{grafvf1}
\end{figure}

In the low-velocity regime $v_{\textrm{F}}\ll 1$, the lowest order term in Eq.~(\ref{runningvf}) reads
\begin{equation}\label{vfmenor}
\beta_{v_{\textrm{F}}}=-\frac{e^{2}}{16\pi(1+\theta^{2})}.
\end{equation}
By solving Eq.~(\ref{vfmenor}), we find the renormalized Fermi velocity, namely,
\begin{equation}\label{vfrstaticregime}
v^{\textrm{R}}_{\textrm{F}}(\mu)=v_{\textrm{F}}(\mu_0)\left[ 1-\frac{\alpha_{\textrm{eff}}}{4}\ln\left(\frac{\mu}{\mu_{0}}\right)\right],
\end{equation}
where
\begin{equation}
\alpha_{\textrm{eff}}=\frac{e^2}{4\pi v_{\textrm{F}}(\mu_0)(1+\theta^2)}.
\end{equation}
From Eq.~(\ref{vfrstaticregime}) is clear that for this theory, the Fermi velocity renormalization is controlled by Chern-Simons parameter. For $\theta$ equal zero, the $v^{\textrm{R}}_{\textrm{F}}$ is the well-known result, caculated in Ref.~\cite{vozmediano}. Nevertheless, for large values of $\theta$, the Fermi velocity does not renormalize, remaning close to its bare value, see Fig.~\ref{grafvf1}


In the high-velocity regime $v_{\textrm{F}}\sim 1$, until the first order in ($1-v_{\textrm{F}}$), $\beta_{v_{\textrm{F}}}$ becomes
\begin{equation}
\beta_{v_{\textrm{F}}}=-\frac{2e^{2}}{5\pi^{2}(1+\theta^{2})}(1-v_{\textrm{F}}).
\end{equation}
Hence, the Fermi velocity for $v_{\textrm{F}}\sim 1$ is
\begin{equation}\label{vfc}
v^{\textrm{R}}_{\textrm{F}}(\mu)=\left[ 1-(1-v_{\textrm{F}}(\mu_0))\left(\frac{n}{n_{0}}\right)^{\frac{\gamma}{2}}\right],
\end{equation}
where $\gamma = \frac{8v_{\textrm{F}}\alpha_{\textrm{eff}}}{5\pi}$ and we have  used the relation $ \mu / \mu_{0} \rightarrow (n/n_{0})^{1/2}$ \cite{Gorbachev}, with $n$ and $n_{0}$ being the density of states.


\begin{figure}[hbtp]
\centering
\hspace{-.25cm}\includegraphics[scale=0.7]{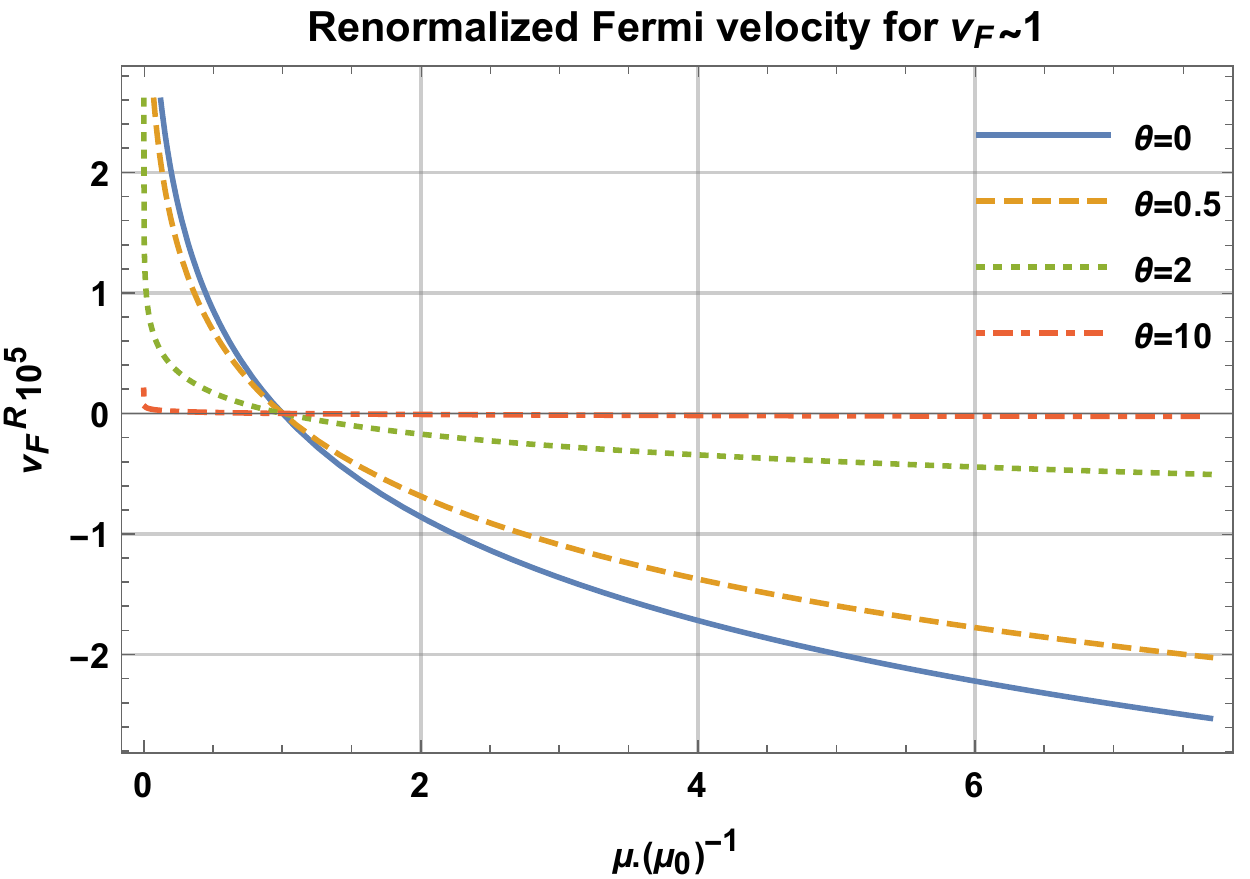}
\caption{\textit{The renormalized Fermi velocity for $v_{\textrm{F}}\sim 1$}. This plot shows the behavior of the renormalized Fermi velocity Eq.~(\ref{vfc}) as a function of $\mu/\mu_{0}$, varying the parameter $\theta$, where the solid line (blue), dashed (orange), dotted (green) and dot-dashed (red) are for $\theta=0$, $\theta=0.5$, $\theta=2$ and $\theta=10$, respectively.}
\label{grafvf}
\end{figure}


In the limit of large values of $\theta$ the Fermi velocity does not renormalize, see Fig.~\ref{grafvf}. Note also that for $n\rightarrow 0$ \cite{marinolivro}, the low-density regime, the Fermi velocity renormalize to $c=1$. This result implies that the renormalized Fermi velocity does not diverge at zero doping, instead of the corresponding result in Eq.~(\ref{vfrstaticregime}).

\section{Discussion}

Planar-field theories are the ideal platform for investigating electronic interactions in two-dimensional materials at low energies. The well-developed methods of quantum field theory for calculating quantum corrections, dynamical symmetry breaking, and anomalies make this idea more exciting yet. However, the full description of these systems is a hard task, mainly because of several kinds of interactions that emerge due to the lattice vibrations, disorder, and impurities just to cite a few. Aiming to this direction, PQED has been investigated together with a Gross-Neveu interaction, which is expected to describe either phonons or impurities at low energies \cite{CSBPQED}. Beyond all of these microscopic interactions, we also have the challenge of describing the topological effects driven by the Chern-Simons action. 

In the present work we have shown that, after including the dynamics of the Chern-Simons action in PQED, the $\theta$-parameter becomes dimensionless and most of its effects may be absorbed by a simple scaling of the electric charge at the static limit. Therefore, our model provides any dielectric constant $\epsilon$ depending on the value of $\theta$, i.e, we may always write $\epsilon=1+\theta^2$ for fitting $\epsilon$. The different values of $\epsilon$ may be obtained experimentally either by placing the material above some substrate or by changing the doping $n$ of the material. Because $\epsilon$ is a function of $n$ \cite{gfactor}, hence, we may assume that our $\theta$-parameter only encodes the effects of $n$, i.e, $\epsilon(n)=1+\theta^2(n)$. This is likely to be a relevant effective description for such effect. Furthermore, within the dynamical limit, the $\theta$-parameter changes the renormalized electron mass $m_R$, yielding a maximal value at $\theta\approx 0.36$. It is worth to mention that this renormalized mass (also called band gap) has been shown to be dependent on the doping for a 2D-material, such as the transition metal dichalcogenide monolayers \cite{gapcontrol}. Finally, we also have calculated the renormalized Fermi velocity in both static and dynamical regimes. 

It should be noted that the normalization of Fermi velocity does not depend if $\bar{A}_{\mu}\neq {\cal A}_{\mu}$ or $\bar{A}_{\mu}= {\cal A}_{\mu}$. In fact, according to Eq.(\ref{2FM}) (or Eq.(\ref{Sigma})) the CS term contribution to the electron self-energy is finite,  therefore, this term does not change the  $v_{\textrm{F}}$ renormalization.

All of these results are dependent on the one-loop perturbation theory, but our model is likely to provide new physics in other regimes, for instance, in the strong-coupling limit or when fermions are coupled to an external magnetic field. These limits are relevant for investigating a dynamical symmetry breaking and fractional quantum Hall effect \cite{FQHE}, respectively. Dynamical effects of PQED, related to topological phases and dynamical mass generation, have been calculated in Ref.~\cite{PRX2015}. It is relevant to understand whether the screening effects may either induce or destroy such phase transitions.

\section*{Acknowledgments}

G. C. M.  is partially supported by Coordenação de Aperfeiçoamento de Pessoal de Nível Superior – Brasil (CAPES), finance code 001; V. S. A. and L. O. N. are partially supported by Conselho Nacional de Desenvolvimento Científico e Tecnológico (CNPq) and by CAPES/NUFFIC, finance code 0112; E. C. M. is partially supported by both CNPq and Fundação Carlos Chagas Filho de Amparo à Pesquisa do Estado do Rio de Janeiro (FAPERJ).

\textbf{\section*{\textbf{Appendix A: Maxwell-Chern-Simons Action}}}

In this appendix we calculate the effective action for the matter current coupled to the MCS gauge field. Thereafter, we investigate the static limit of this model. We start with the action, given by 
\begin{eqnarray}
{\cal L}_{\rm{QED}}&=&\frac{1}{4}\tilde F^{\mu\nu} \tilde F_{\mu\nu}+\lambda \tilde A^\mu \partial_\mu \partial_\nu \tilde A^\nu \nonumber\\
&+&{\cal L}_M[\psi]+i \frac{\theta}{2} \epsilon_{\mu\nu\alpha}\tilde A^\mu \partial^\nu \tilde A^\alpha+e j^\mu \tilde A_\mu,  \label{QEDCS}
\end{eqnarray}
where $\tilde A_{\mu}$ is the MCS field \cite{Kondo,anomaly}. Integration over $ \tilde A_{\mu}$ yields
\begin{eqnarray}
{\cal L}^{\rm{QED}}_{\rm{eff.}}[j_\mu]&=&-\frac{e^2}{2}j^\mu \left[\frac{1}{-\Box+4\theta^2}\right]j_\mu\nonumber \\
&+&i\frac{e^2\theta}{2}j^\mu\epsilon_{\mu\nu\alpha}\partial^\alpha\left[\frac{1}{\Box(-\Box+4\theta^2)}\right] j^\nu. \label{qedeff}
\end{eqnarray}

It is not surprising that the first term in the rhs of Eq.~(\ref{qedeff}) does not provide the proper electromagnetic effective action, which is given by PQED. The effect on the statistical interaction, nevertheless, is more dramatic. Indeed, this interaction is proportional to $\Box^{-1}$ in Eq.~(\ref{CS2}). Here, it is modified to $\Box^{-1}(-\Box+4\theta^2)^{-1}$, which yields some different kind of interaction.  This occurs because $\theta$ has the dimension of mass in Eq.~(\ref{QEDCS}). For PQED coupled to the Chern-Simons action, nevertheless, $\theta$ is dimensionless. We may estimate the effect of a massive $\theta$ by calculating the static limit. In this case, the static potential generated by the model in Eq.~(\ref{QEDCS}) is
\begin{equation}
V(r)=\frac{e^2}{2\pi} K_0(r \theta),
\end{equation}
where $K_0(r\theta)$ is the modified Bessel function of the second kind. For $r\theta\ll 1$, we have $V(r)\propto \ln(r \theta)$, while for $r\theta\gg 1$, we find $V(r)\propto e^{-r\theta}/\sqrt{r\theta}$. It is clear that $1/\theta$ plays the role of an interaction length, wherein the short-range limit we have a logarithmic-confining potential. On the other hand, in the long-range limit, we find an exponential potential which quickly goes to zero. This resembles the Meissner effect in superconductors.

The main goal of this appendix is to clarify the relevance of our model in Eq.~(\ref{action}). Indeed, it is remarkable that PQED coupled to the Chern-Simons action provides a simultaneous description for both electromagnetic and statistical interactions in the light of the effective action for the matter current.

\textbf{\section*{\textbf{Appendix B: Mass renormalization}}}

In this appendix we derive Eq.~(\ref{mrm-1}) that yields the physical mass. This renormalized mass $m_{\textrm{R}}$ is given by the pole of the corrected propagator, given by
\begin{equation}\label{fermionpropc}
S^{\textrm{R}}_{F}=\frac{-1}{\slashed{p}-m-\Sigma^{\textrm{R}}(\slashed{p},m)}.
\end{equation}
Using Eq.~(\ref{sigR}) in Eq.~(\ref{fermionpropc}), we find
\begin{equation}\label{fermionpropc2}
S^{\textrm{R}}_{F}=\frac{-1}{(1-C)\slashed{p}-(1+D)m},
\end{equation}
where $C$ and $D$ are functions of $p$ and $m$ given by Eq.~(\ref{C}) and Eq.~(\ref{D}), respectively. Furthermore, they are proportional to $e^{2}$ within our one-loop approximation. From Eq.~(\ref{fermionpropc2}), we conclude that the pole of the propagator reads $p^2=m^2(1+D)^{2}/(1-C)^2$. Within the one-loop approximation, this can be rewritten as
$p^2=M^2(p,m)$, where $M(p,m)=1+C+D$ is called mass function. 
Therefore, the renormalized mass, in the mass shell, is given by
\begin{equation}
m_{\textrm{R}}=M(p^{2}=-m^{2})=1+C+D,
\end{equation} 
which is our desired result in Eq.~(\ref{mrm-1}). Note that to find the renormalized mass it is necessary to return to Minkowski space, for more details regarding this approximation see \cite{Nascimento}.

\textbf{\section*{\textbf{Appendix C: Renormalization Group Functions}}}

The vertex functions, which have been made finite by the subtraction of the pole terms in the dimensionally regularized amplitudes, satisfy a 't Hooft-Weinberg renormalization group equation \cite{Weinberg}:
\begin{equation}
\Big( \mu\frac{\partial}{\partial\mu}+\beta_e\frac{\partial}{\partial e}+\beta_{v_{\textrm{F}}}\frac{\partial}{\partial v_{F}}-N_F\gamma_{\psi}-N_A\gamma_A \Big) \Gamma^{(N_F,N_A)}=0,
\end{equation}
where $\Gamma^{(N_F,N_A)}=\Gamma^{(N_F,N_A)}(p_i,\mu ,e,v_{\textrm{F}})$ denotes the renormalized vertex function of $N_F$ fermion fields and $N_A$ gauge fields, and $p_i$ denotes the external momenta. We have introduced the renormalization scale $\mu$ through the substitution $e\rightarrow \mu^{\epsilon /2}\,e$, where $\epsilon =2-d$. The $\beta_i$'s for $i=e,v_{\textrm{F}}$ are the beta functions which describe how the electric charge, Fermi velocity change with the scale parameter $\mu$ and are usually defined as $\beta_e=\mu \frac{\partial e}{\partial\mu}$, $\beta_{v_{\textrm{F}}}=\mu \frac{\partial v_{\textrm{F}}}{\partial\mu}$.

The functions $\gamma_{\psi}$ and $\gamma_A$ are the anomalous dimensions of the fields $\psi$ and $A_{\mu}$ and given by $\gamma_{\psi}=\mu \frac{\partial}{\partial\mu}(\ln \sqrt{Z_{\psi}})$ and $\gamma_{A}=\mu \frac{\partial}{\partial\mu}(\ln \sqrt{Z_{A}})$, where $Z_{\psi}$ and $Z_A$ are the wave function renormalization of the fields $\psi$ and $A_{\mu}$, respectively. 

To remove the pole term in the amplitudes $I^{(N_F,N_A)}$ contained in $\Gamma^{(N_F,N_A)}$, we use the following prescription 
\[
(1-{\cal T})\mu^{x\epsilon}\,I^{(N_F,N_A)}=\mbox{Finite}^{(N_F,N_A)}+x\ln\mu\,\mbox{Res}^{(N_F,N_A)},
\]
where ${\cal T}$ is an operator used to remove the pole term. In the above expression, $\mbox{Res}^{(N_F,N_A)}$ means residue of the diagram that is given by the coefficient of the term $1/\epsilon$, and $\mbox{Finite}^{(N_F,N_A)}$ is the finite part of the amplitudes $I^{(N_F,N_A)}$.

By performing the calculations up to one loop, we find
\begin{equation}
\Gamma^{(2,0)}(p)= \gamma^{0}p_{0}+v_{\textrm{F}}\gamma^{i}p_{i} +\Sigma (p) \ ,
\end{equation}
where $\Sigma (p) = \left[\mbox{Finite}^{(2,0)}+\ln\mu\,\mbox{Res}^{(2,0)}\right]$, and $\mbox{Res}^{(2,0)}=A_1\,\gamma^0p_0+A_2\,\gamma^i p_i$. 

After writing the RG functions perturbatively as, $\beta_i=\sum_{j=1}^{2}\beta_{i}^{(j)}\,e^j$, we find that in order of $e$, 
\begin{eqnarray}
\gamma_{\psi}^{(1)}=0, \qquad \beta_{v_{\textrm{F}}}^{(1)}= 0, \qquad \beta_{e}^{(1)}=0, \ . \nonumber 
\end{eqnarray}

On the other hand, in order $e^2$ we find
\begin{eqnarray}
\gamma_{\psi}^{(2)}&=&\frac{1}{2}\,A_1, \qquad \beta_{v_{\textrm{F}}}^{(2)}= A_2-v_{\textrm{F}}\,A_1, \nonumber \\
 \beta_{e}^{(2)}&=&\gamma^{(1)}_A,\ . \nonumber 
\end{eqnarray}

However, because the pholarization tensor is finite in one-loop (order $e^2$) and accordingly $\beta_e=0$.

Finally, we may identify from the Feynman diagrams that
\begin{eqnarray}
A_1 &=& -\frac{e^2}{8\pi^2(1+\theta^2)}(1-2v_{\textrm{F}}^2)\,I_1, \nonumber\\
A_2 &=& \frac{e^2}{8\pi^2 (1+\theta^2)}v_{\textrm{F}} I_3,\nonumber
\end{eqnarray}
and we obtain
\begin{equation}\label{betavf}
\beta_{v_{\textrm{F}}}=\frac{e^2}{8\pi^2 (1+\theta^2)}\,\left[v_{\textrm{F}}(1-2v_{\textrm{F}}^2)\,I_1+v_{\textrm{F}}\,I_3\right].
\end{equation}

The flow of the effective Fermi velocity can be written as:
\begin{equation}
\frac{\partial v^R_F (\mu) }{\partial t}=\beta_{v_{\textrm{F}}}(\beta), \label{defvrenor}
\end{equation}
where $v_{\textrm{F}}(t=0)=v_{\textrm{F}}(\mu_0)$,  and we have introduced a logarithmic scale $t=\ln(\mu/\mu_0)$, with $\mu_0$ a reference scale where the parameter $v_{\textrm{F}}^0$ has been defined.

By solving Eq.(\ref{defvrenor}) for $v_{\textrm{F}}\ll 1$ and $v_{\textrm{F}}\sim 1$ regime we get the Eqs.(\ref{vfrstaticregime}) and (\ref{vfc}).

\clearpage

\end{document}